\documentclass[american]{article}
\usepackage[T1]{fontenc}
\usepackage[latin9]{inputenc}
\usepackage{amsmath}
\usepackage{graphicx}
\usepackage{esint}
\usepackage[numbers]{natbib}

\makeatletter
\usepackage{mciteplus}

\makeatother

\usepackage{babel}

\begin{document}

\title{Off-shell helicity amplitudes in high-energy factorization\thanks{Presented at  the Low x workshop, May 30 - June 4 2013, Rehovot and Eilat, Israel}}

\author{Piotr Kotko\footnote{piotr.kotko@ifj.edu.pl}\, \thanks{In collaboration with A. van Hameren and K. Kutak}\vspace{10pt} \\ \textit{\small The H.\ Niewodnicza\'nski Institute of Nuclear Physics} \\ \textit{\small Polish Academy of Sciences}\\ \textit{\small Radzikowskiego 152, 31-342 Cracow, Poland}}

\date{{}}
\maketitle
\begin{abstract}
In the Catani-Ciafaloni-Hautmann high-energy factorization approach
a cross section is expressed as a convolution of unintegrated gluon
densities and a gauge-invariant hard process, in which two incoming
gluons are off-shell with momenta satisfying certain high-energy kinematics.
We present two methods of evaluating the tree-level hard process with
multiple final states. The first one assumes that only one of the
gluons is off-shell and relies on the Slavnov-Taylor identities. Such
asymmetric configuration of incoming gluons is phenomenologically
important in small x probing by forward processes. The second method
deals also with two off-shell gluons and is based on the analytic
continuation of the off-shell gluons momenta to the complex space.
The methods were implemented into Monte Carlo computer programs and
used in phenomenological applications. The results of both methods
are straightforwardly related to Lipatov's effective vertices in quasi-multi-regge
kinematics.
\end{abstract}

\section{Introduction}

It is commonly known that in small x physics one needs a resummation
of certain types of logarithms, that otherwise spoil the calculation.
Such a procedure is provided by so-called \textit{high-energy factorization}
of Catani, Ciafaloni and Hautmann (CCH) \citep{Catani:1990eg,Catani:1990ka,Catani:1990xk,Catani:1994sq}.
Its main ingredients are unintegrated gluon densities and \textit{off-shell
matrix elements}, which are the subject of the present talk. In general,
only in simple cases an ordinary high-energy amplitude (i.e. evaluated
from standard Feynman diagrams) can be gauge invariant despite its
off-shellness. The CCH factorization was thus originally constructed
for heavy quark production which is the exception just mentioned.
At present, it is however highly desirable to use this approach to
more complicated partonic final states. 

One of the methods providing gauge invariant matrix elements that
suits CCH approach is given in terms of the Lipatov's effective action
\citep{Lipatov:1995pn}. The resulting Feynman rules \citep{Antonov:2004hh}
involve (except standard QCD rules) additional so called induced vertices
which possess complicated structure. Indeed, subsequent vertices with
increasing number of legs have to be constructed recursively and explicit
constructions exist up to five external legs \citep{Antonov:2004hh}.
For the recent applications of that approach in LHC phenomenology
see e.g. \citep{Nefedov:2013ywa,Saleev:2012np,Kniehl:2011hc,Hentschinski:2013bp}.

Let us point out, that the amplitudes we are talking about are merely
tree-level amplitudes. In collinear approach (utilizing on-shell amplitudes)
there are presently a lot of tools and methods allowing for automatic
and efficient calculation of any tree-level process. This is certainly
not the case for the off-shell amplitudes. Therefore two new method
have been proposed in Refs. \citep{vanHameren:2012if,vanHameren:2012uj}.
The second method applies in certain, yet very important phenomenologically,
situation of forward jet production processes. In what follows, after
recalling the framework of high-energy factorization, we shall briefly
describe the both methods.

\section{High-energy factorization}

\label{sec:Factoriz}

\begin{figure}
\begin{centering}
\parbox{5cm}{A)\\\\\includegraphics[width=5cm]{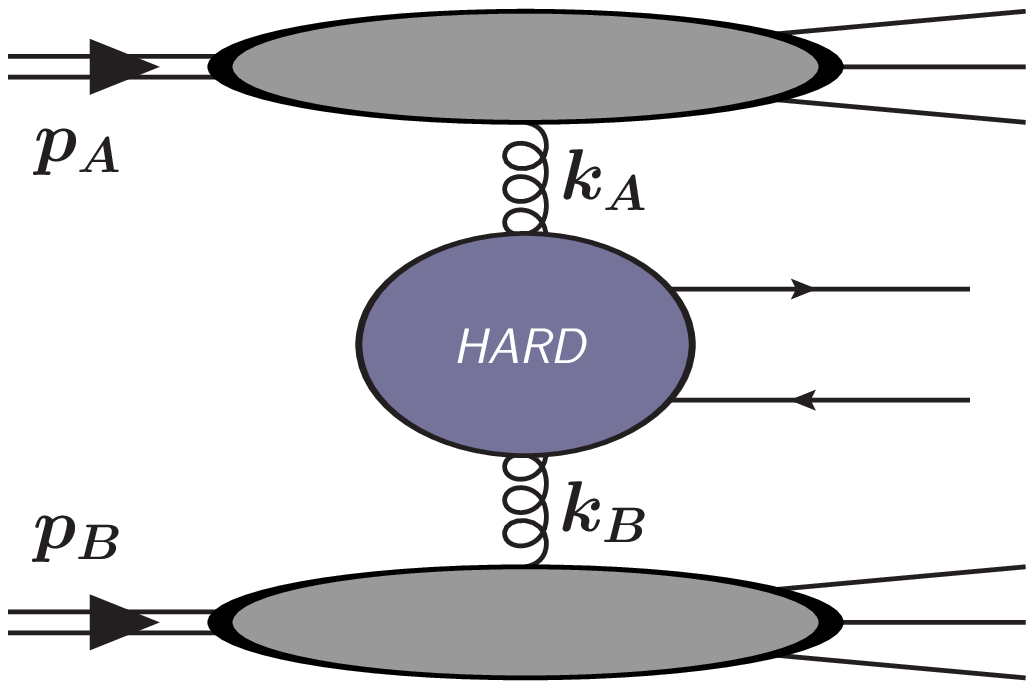}}~~~~~~~~~~~~\parbox{5cm}{B)\\\\\includegraphics[width=5cm]{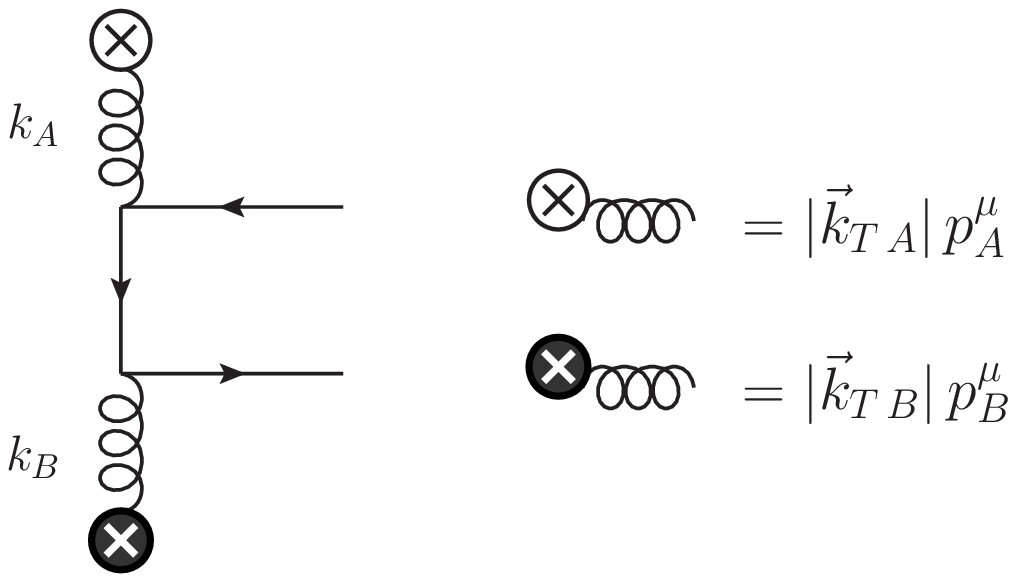}}
\par\end{centering}

\caption{\label{fig:factoriz}\small A) High-energy factorization for $p_{A}p_{B}\rightarrow Q\overline{Q}$
into unintegrated PDFs (top and bottom blobs after 'squaring') and
parton-level sub-process $g^{*}g^{*}\rightarrow Q\overline{Q}$ (middle
blob). B) The hard sub-process is defined by an off-shell matrix element
with incoming off-shell gluon propagators contracted with high-energy
projectors (explained on the r.h.s). In this particular case the amplitude
is gauge invariant.}
\end{figure}

The basic statement of CCH factorization is that at high energies,
the cross section for the process for heavy quark hadroproduction
can be expressed as \begin{multline}
d\sigma_{AB\rightarrow Q\overline{Q}}=\int d^{2}k_{T\, A}\int\frac{dx_{A}}{x_{A}}\,\int d^{2}k_{T\, B}\int\frac{dx_{B}}{x_{B}}\,\\
\mathcal{F}_{g^{*}/A}\left(x_{A},k_{T\, A}\right)\,\mathcal{F}_{g^{*}/B}\left(x_{B},k_{T\, B}\right)\, d\hat{\sigma}_{g^{*}g^{*}\rightarrow Q\overline{Q}}\left(x_{A},x_{B},k_{T\, A},k_{T\, B}\right),\label{eq:HEN_fact_1}\end{multline}
where $A$, $B$ are hadrons with momenta $p_{A}$, $p_{B}$ respectively,
$\mathcal{F}_{g^{*}/H}\left(x,k_{T}\right)$ are unintegrated gluon
densities undergoing the BFKL evolution and $d\hat{\sigma}_{g^{*}g^{*}\rightarrow Q\overline{Q}}$
is the hard cross section for the process $g^{*}g^{*}\rightarrow Q\overline{Q}$
at tree level (Fig \ref{fig:factoriz}A ). The momenta of the off-shell
gluons $g^{*}$ have the following high-energy form\begin{equation}
k_{A}^{\mu}=x_{A}p_{A}^{\mu}+k_{T\, A}^{\mu},\,\,\,\,\, k_{B}^{\mu}=x_{B}p_{B}^{\mu}+k_{T\, B}^{\mu},\label{eq:kAkB}\end{equation}
where $p_{A}\cdot k_{T\, A,B}=p_{B}\cdot k_{T\, A,B}=0$. The amplitude
for the process $g^{*}g^{*}\rightarrow Q\overline{Q}$ is constructed
using the ordinary diagram retaining however the gluon propagators
and contracting them with eikonal vertices defined as $\left|\vec{k}_{T\, A}\right|p_{A}^{\mu}$
and $\left|\vec{k}_{T\, B}\right|p_{B}^{\mu}$ (Fig. \ref{fig:factoriz}B
). The amplitude is gauge invariant, fundamentally due to the form
of the projectors.

In what follows we assume that the above factorization is still valid%
\footnote{In the present short article we put aside comments about validity
of $k_{T}$-factorization at small x. We have gathered together some
basic facts in \citep{vanHameren:2013fla}, so we refer the reader
to that paper and references therein.%
} when replacing $Q\overline{Q}$ by any partonic state $X$. The problem
is, however, that simple generalization of the above prescription
to calculate the hard matrix element does not lead to the gauge invariant
result. As already mentioned in the Introduction additional contributions
are needed. A general method to overcome this difficulty will be described
in Section \ref{sec:twoofshell}.

Let's suppose now that we are interested in a situation where $x_{A}\ll x_{B}$
which occurs typically when one tries to access small x by looking
at the forward jets. Since $x_{B}$ is large, one may assume that
the gluon with momentum $k_{B}$ is nearly on-shell and transform
the Eq. (\ref{eq:HEN_fact_1}) into\begin{multline}
d\sigma_{AB\rightarrow X}=\int d^{2}k_{T\, A}\int\frac{dx_{A}}{x_{A}}\,\int dx_{B}\,\\
\sum_{b}\mathcal{F}_{g^{*}/A}\left(x_{A},k_{T\, A}\right)\, f_{b/B}\left(x_{B}\right)\, d\hat{\sigma}_{g^{*}b\rightarrow X}\left(x_{A},x_{B},k_{T\, A}\right),\label{eq:HENfact_2}\end{multline}
where $b$ runs over gluon and all the quarks that can contribute
to the production of multiparticle state $X$ (see also \citep{Deak:2009xt}).
Now the hard process has a single leg off-shell, what somewhat simplifies
the situation. A suitable method of evaluating such amplitudes will
be outlined in Section \ref{sec:oneofshell}.

\section{The general method}

\label{sec:twoofshell}

Let us start with pointing out what are the features that we require
from the new approach. First, it should use helicity method, which
-- generally speaking -- is based on utilizing helicity spinors as
a basic object used to construct the amplitudes. Second, it should
be easily implementable in efficient computer programs, similar to
the tools like HELAC for example \citep{Cafarella:2007pc}.

\begin{figure}
\begin{centering}
\includegraphics[width=9cm]{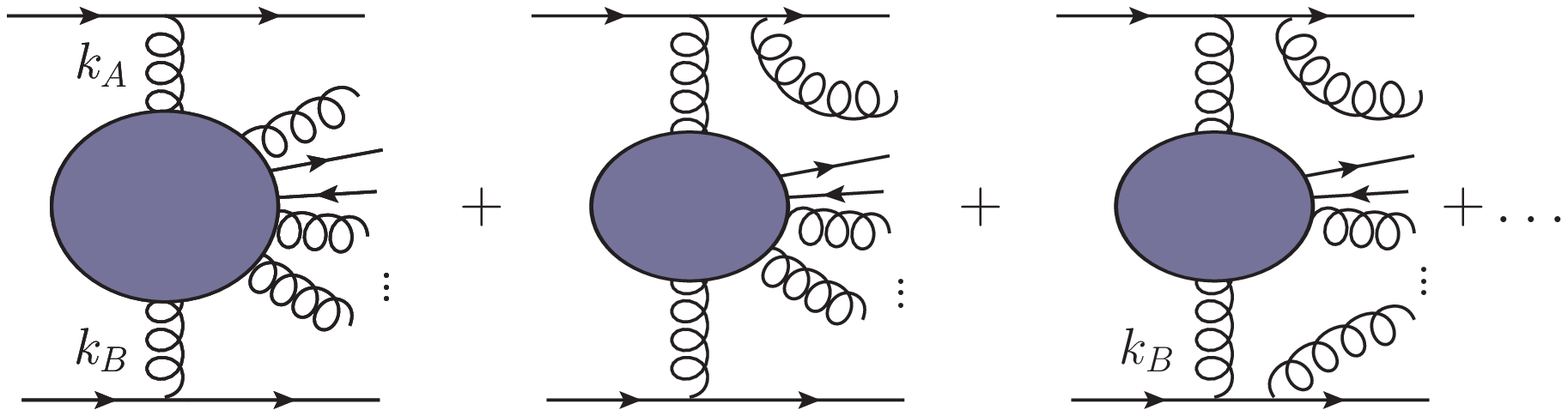}
\par\end{centering}

\caption{\label{fig:twooffshell}\small Embedding the off-shell process $g^{*}g^{*}\rightarrow X$
into a process with external quarks. It is impossible to have the
exact high-energy kinematics of the momenta transfers $k_{A}$, $k_{B}$
to the $g^{*}g^{*}\rightarrow X$ subprocess and keep all the quarks
on-shell. The solution is to allow for complex momenta of the external
quarks.}

\end{figure}

A rather obvious attempt towards such an approach would be to embed
the off-shell process $g^{*}g^{*}\rightarrow X$ into a bigger gauge
invariant process $q\left(p_{A}\right)q\left(p_{B}\right)\rightarrow q'\left(p'_{A}\right)q'\left(p'_{B}\right)X$
as depicted in Fig. \ref{fig:twooffshell}. It is however easy to
see that it is impossible to maintain the on-shellness of all of the
external partons and keep the momenta transfers $p'_{A}-p_{A}=k_{A}$
and $p'_{B}-p_{B}=k_{B}$ in the form of (\ref{eq:kAkB}) in the same
time. Obviously, compromising on-shellness of any of the quarks would
spoil the gauge invariance. We may, however, as pointed out in \citep{vanHameren:2012if},
compromise real-valuedness of the momenta of the quarks. This will
of course make the whole amplitude non-physical; the point is however
that we are interested not in the quark amplitude, but rather in the
off-shell amplitude.

To illustrate the method in some more details let us introduce four
basis null vectors: real-valued $l_{1}$, $l_{2}$ and complex-valued
$l_{3}$, $l_{4}$ defined as\begin{equation}
l_{3}^{\,\mu}=\frac{1}{2}\left\langle l_{2};-\right|\gamma^{\mu}\left|l_{1};-\right\rangle ,\,\,\,\,\,\, l_{4}^{\,\mu}=\frac{1}{2}\left\langle l_{1};-\right|\gamma^{\mu}\left|l_{2};-\right\rangle ,\end{equation}
where $\left|l;\pm\right\rangle $ is a massles spinor corresponding
to momentum $l$ and helicity $\pm$ (for more detail about helicity
formalism see e.g. \citep{Mangano:1990by}). They satisfy $l_{1,2}\cdot l_{3,4}=0$
and $l_{1}\cdot l_{2}=-l_{3}\cdot l_{4}$. The complex vectors $l_{3}$,
$l_{4}$ play the role of transverse vectors. Using this basis one
may decompose the external quarks momenta as follows\begin{gather}
p_{A}^{\,\mu}=\left(\Lambda+x_{A}\right)l_{1}^{\,\mu}-\frac{l_{4}\cdot k_{T\, A}}{l_{1}\cdot l_{2}}\, l_{3}^{\,\mu},\,\,\,\,\,\,\,\, p_{B}^{\,\mu}=\left(\Lambda+x_{B}\right)l_{2}^{\,\mu}-\frac{l_{3}\cdot k_{T\, B}}{l_{1}\cdot l_{2}}\, l_{4}^{\,\mu},\\
p'{}_{A}^{\,\mu}=\Lambda l_{1}^{\,\mu}+\frac{l_{3}\cdot k_{T\, A}}{l_{1}\cdot l_{2}}\, l_{4}^{\,\mu},\,\,\,\,\,\,\,\, p'{}_{B}^{\,\mu}=\Lambda l_{2}^{\,\mu}+\frac{l_{4}\cdot k_{T\, B}}{l_{1}\cdot l_{2}}\, l_{3}^{\,\mu},\end{gather}
where $\Lambda$ is a real parameter. Note, that this decomposition
preserves both on-shellness $p_{A,B}^{2}=p'\,_{A,B}^{2}=0$ and high-energy
kinematics (\ref{eq:kAkB}) for any $\Lambda$. Another important
property is that the spinors for external quarks satisfy the following
proportionality relations $\left|p_{A};-\right\rangle \propto\left|l_{1};-\right\rangle $,
$\left|p_{B};-\right\rangle \propto\left|l_{2};-\right\rangle $.
Thus, we may trade the original spinors to the longitudinal spinors
without spoiling the gauge invariance and thus simplifying the calculation.
In order to decouple the unphysical (basically complex) degrees of
freedom one has to take the smooth limit $\Lambda\rightarrow\infty$.
In principle it can be done numerically, however the better solution
is to do it analytically. Only external quark lines are directly affected
by this limit and they turn out to be reduced to eikonal couplings
and propagators, however in a fully controlled manner. The method
was implemented in a MC code similar to HELAC and used to calculate
certain distributions with four and five partons in the final state
\citep{vanHameren:2012if}, what demonstrates its power.

\section{One-leg off-shell amplitudes}

\label{sec:oneofshell}

As mentioned in Section \ref{sec:Factoriz}, in small x practice one
mostly needs the high-energy amplitudes with just a single gluon being
off-shell. Of course the method described in the previous section
applies here as well. Nevertheless, there is another interesting method
\citep{vanHameren:2012uj} (predating the former) which we are now
going to outline.

\begin{figure}
\begin{centering}
\parbox{4cm}{A)\\\\\includegraphics[width=2.5cm]{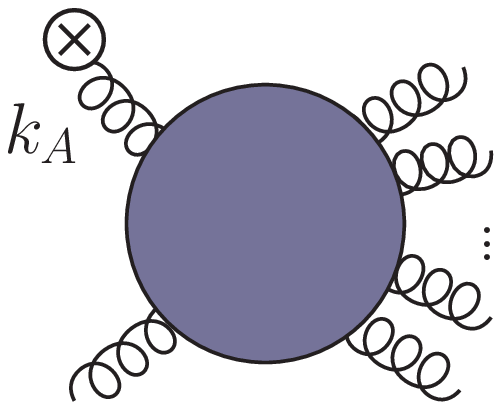}}~~~~~~~\parbox{4cm}{B)\\\\\includegraphics[width=3.5cm]{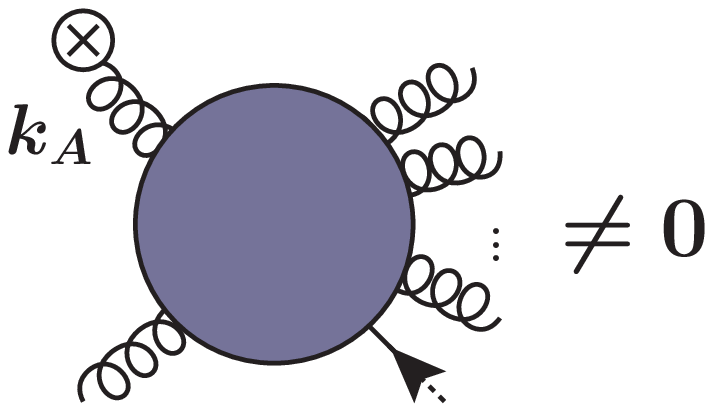}}
\par\end{centering}

\caption{\label{fig:singleoffshell1}\small A) Multigluon amplitude with a
single leg off-shell. B) Violation of the Ward identity -- replacement
of any of the polarization vectors by the momentum (indicated by an
arrow) leads to a non-zero result.}
\end{figure}

First, let us point out, that the CCH factorization is formulated
in axial gauge. Thus, as a gauge invariant amplitudes we mean the
ones that satisfy the Ward identities. To be more specific, let us
denote the amplitude with off-shell leg $k_{A}$ and $N$ final state
gluons as $\mathcal{M}\left(\varepsilon_{1},\ldots,\varepsilon_{N}\right)$
where $\varepsilon_{i}$ are polarization vectors (Fig. \ref{fig:singleoffshell1}A
). For a general choice of $\varepsilon_{i}$ the amplitude does not
satisfy the Ward identity (Fig. \ref{fig:singleoffshell1}B ) \begin{equation}
\mathcal{M}\left(\varepsilon_{1},\ldots,k_{i},\ldots,\varepsilon_{N}\right)\neq0.\label{eq:Wardviol}\end{equation}
The question that one can ask, is what is the actual value on the
r.h.s. of (\ref{eq:Wardviol}) and whether one can use that information
to construct a new amplitude\begin{equation}
\tilde{\mathcal{M}}\left(\varepsilon_{1},\ldots,\varepsilon_{N}\right)=\mathcal{M}\left(\varepsilon_{1},\ldots,\varepsilon_{N}\right)+\mathcal{W}\left(\varepsilon_{1},\ldots,\varepsilon_{N}\right)\label{eq:Mtild}\end{equation}
such that \begin{equation}
\tilde{\mathcal{M}}\left(\varepsilon_{1},\ldots,k_{i},\ldots,\varepsilon_{N}\right)=0.\end{equation}

\begin{figure}
\begin{centering}
\parbox{11cm}{A)\\\\\includegraphics[width=11cm]{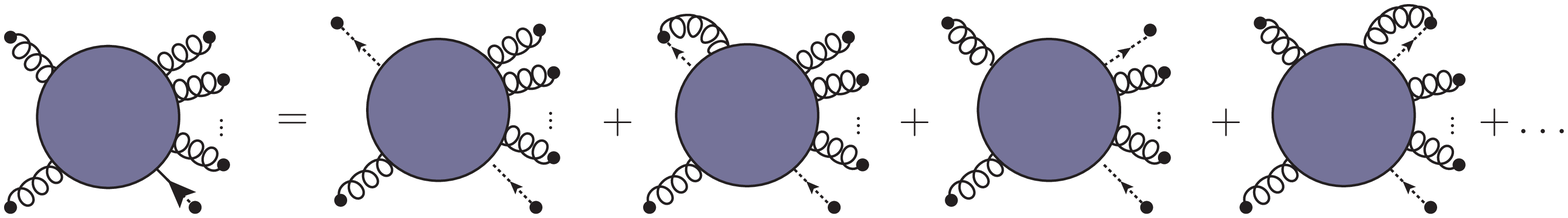}}\bigskip{}

\par\end{centering}

\begin{centering}
\parbox{11cm}{B)\center{\includegraphics[width=4.5cm]{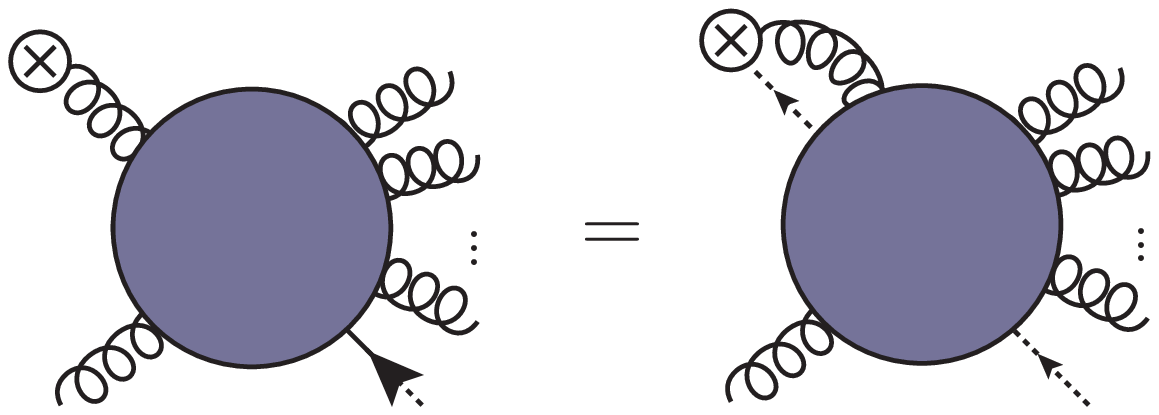}}}
\par\end{centering}

\caption{\label{fig:singleoffshell2}\small A) The Slavnov-Taylor identity
applied to the Green's function. B) After applying the high-energy
reduction formula the single term is left, which is precisely the
amount of gauge-invariance violation.}
\end{figure}

The solution is provided by the basics of QCD, namely by the Slavnov-Taylor
identities (see e.g. \citep{Arodz:2010} for an elementary review).
They however operate on the Green's function level, therefore we need
some sort of a reduction formula for high-energy factorization. It
can be naturally written as\begin{multline}
\mathcal{M}\left(\varepsilon_{1},\ldots,\varepsilon_{N}\right)=\lim_{k_{A}\cdot p_{A}\rightarrow0}\,\lim_{k_{1}^{2}\rightarrow0}\ldots\lim_{k_{N}^{2}\rightarrow0}\\
\left|\vec{k}_{T\, A}\right|p_{A}^{\mu_{A}}\,\, k_{1}^{2}\varepsilon_{1}^{\mu_{1}}\ldots\, k_{N}^{2}\varepsilon_{N}^{\mu_{N}}\,\,\tilde{G}_{\mu_{A}\mu_{1}\ldots\mu_{N}}\left(k_{A},k_{1},\ldots,k_{N}\right),\label{eq:reduction}\end{multline}
where $\tilde{G}$ is the momentum space Green's function. The contraction
of an external leg of $\tilde{G}$ with the corresponding momentum
leads to gauge terms with ghost lines%
\footnote{The fact that we use axial gauge for internal lines does not interfere
with the usage of ghosts. Ghosts can be introduced in axial gauge,
but they decouple from on-shell processes.%
} (Fig. \ref{fig:singleoffshell2}A ). After applying the reduction
formula (\ref{eq:reduction}) the single term survives, which is precisely
the r.h.s. of (\ref{eq:Wardviol}) (Fig. \ref{fig:singleoffshell2}B
). Further, it turns out that by choosing the axial-gauge vector to
be $p_{A}$ the {}``gauge-restoring amplitude'' $\mathcal{W}$ in
(\ref{eq:Mtild}) can be constructed by summing all the gauge contributions
and trading the external ghosts for the longitudinal projections of
polarization vectors. The result turns out to be very simple\begin{multline}
\mathcal{W}_{12\ldots N}\left(\varepsilon_{1},\ldots,\varepsilon_{N}\right)=-\left(\frac{-g}{\sqrt{2}}\right)^{N}\left|\vec{k}_{T\, A}\right|\,\\
\times\frac{\varepsilon_{1}\cdot p_{A}\ldots\varepsilon_{N}\cdot p_{A}}{k_{1}\cdot p_{A}\,\left(k_{1}-k_{2}\right)\cdot p_{A}\ldots\left(k_{1}-\ldots-k_{N-1}\right)\cdot p_{A}},\label{eq:W}\end{multline}
where the subscript denotes that this result correspond to the specific
color ordering of the external legs (the final answer is the sum of
contributions for all color orderings). The case with quarks is actually
very simple and does not require any {}``gauge-restoring amplitudes'',
as can be seen by analyzing the Slavnov-Taylor identities. The above
result allows for a very simple calculation of the pertinent amplitudes
using the Berends-Giele recursion relations \citep{Berends:1987me}
and \textit{any} polarization vectors. This method was implemented
in the Monte Carlo program $\mathtt{LxJet}$ which uses FOAM algorithm
\citep{Jadach:2002kn}.

\section{Summary}

We have presented the two methods of constructing gauge invariant
off-shell amplitudes relevant to high-energy factorization. They correspond
to Lipatov's vertices in quasi-multi-regge kinematics. We have implemented
the methods in the two independent Monte Carlo codes that allow to
calculate the actual cross sections. Those tools have been recently
\citep{vanHameren:2013fla} used to calculate the cross sections for
three jet production at the LHC in the saturation regime using the
unintegrated gluon densities from \citep{Kutak:2012rf}. The method
of Section \ref{sec:twoofshell} was recently extended in \citep{vanHameren:2013csa}
to include off-shell quarks.

\section*{Acknowledgments}

The author  was supported by the NCBiR grant LIDER/02/35/L-2/10/NCBiR/2011.

\bibliographystyle{apsrevM}
\bibliography{fw3j}

\end{document}